# HAPTIC FEEDBACK IN NATURAL ORIFICE TRANSLUMINAL ENDOSCOPIC SURGERY (NOTES)

[1]**Thanh Nho Do** and [2]**Soo Jay Phee**

*Robotic Research Center, School of Mechanical & Aerospace Engineering, Nanyang Technological University, 50 Nanyang Avenue, Singapore 639798*

Email: [1]<thanh4@e.ntu.edu.sg>; [2]<msjphee@ntu.edu.sg>

## ABSTRACT

Flexible tendon-sheath mechanism is commonly used in natural orifice transluminal endoscopic systems because it offers high flexibility, light weight, and easy transmission. Due to the size constraints and sterilization problems, traditional sensors like force/torque sensor are extremely difficult to place at the tool tips of surgical arms. In addition, nonlinear dynamic friction and backlash hysteresis cause challenges to provide haptic feedback to the surgeons when the robotic arms inside the patient's body. Hence, it is extremely difficult to provide the force information to haptic devices and subsequently to the surgeons. To deal with these problems, in this paper we propose a new approach of friction model in the tendon-sheath mechanism (TSM) to provide the force at distal end of endoscopic system. In comparison with current approaches in the literature, the proposed model is able to provide force information at zero velocity and it is smooth. In addition, the model is independent configuration and able to capture friction force with any complex sheath shapes. A suitable experimental setup is established to validate the proposed approach using the Two-DOFs Master-Slave system. The validity of the proposed approach is confirmed with a good agreement between the estimated model and real experimental data. Finally, a haptic feedback structure is also given for use in flexible endoscopic systems.

## 1. INTRODUCTION

Flexible endoscopy allows for treating and inspecting gastrointestinal (GI) disorders without leaving any abdominal incision to the human's body [1-7]. One of the promising techniques using flexible endoscope is natural orifice transluminal endoscopic surgery (NOTES). It can approach desired targets using flexible endoscope and robotic arms to perform complex surgical task like suturing. It can operate in small and narrow paths, provides high load and more flexibility. A pair of tendon-sheath mechanism (TSM) is often used to actuate the robotic arm. Although the TSM promises beneficial features for transmission, nonlinearities and backlash hysteresis cause loss of tension from the proximal end to distal end of the system. This can be disadvantages in medical applications where traditional sensors like strain gauges cannot be placed at the tool tips of robotic arm due to the size constraint and sterilization problems.

Haptic interface are devices that provide and simulate the sense of touch from the tissues to the surgeon's hand. In medical applications, it is a vital factor because the surgeon is able to feel the differences between soft or hard objects. With the availability of current haptic devices such as CyberForce, CyberGrasp, Phantom, the capability of enhancing safe surgery and performances for surgical systems are always possible [8, 9]. It was reported that haptic feedback to the surgeons would be an essential factor for safe surgery [10-12]. Without such feedback, surgeons are not able to feel as they have in direct contact with the real tissues. Since sensors cannot be mounted at the distal end, the capability of providing force information to haptic devices and subsequently to surgeons is extremely difficult. Therefore, mathematical model can be considered as potential substitutes. Recently, several researchers have discussed and analyzed the transmission problems of the TSM in various applications. Kaneko *et al.* [13-15] presented the tension transmission of the TSM using Coulomb friction model. Do *et al.* [16-18] introduced the displacement transmission for a single TSM. However, no tension analyses have been discussed. Palli *et al.* [19], Tian *et al.* [20], Chen *et al.*[21] and Low *et.al* [22] introduced the transmission for a single TSM with the assumption of the same pretension for small elements. Agrawal *et al.* [23] used a set of partial differential equations to model a single TSM and a pair of TSM in a closed loop approach. Do *et al*. [18, 24-28] introduced compensation control for the TSM. These existing approaches only consider the transmission model for the TSM when the configuration is known and the sheath curve angles are available. In addition, the use of Coulomb model causes discontinuity when the system operates at vicinity of zero velocity. Nonlinear friction in a pair of TSM is quite complex. The forces are different when the system is accelerating and decelerating. Current friction models like Dahl, LuGre, Leuven, or GMS [29, 30] are not able to describe the complete features of nonlinear friction in a pair of TSM since they mainly model the friction force with the same values in acceleration and deceleration using Stribeck curve. Although Do *et al*. addressed these problems in [31-35], they only considered the friction model for a single TSM and the models possess a high number of parameters. It will cost more time to identification and computation. In this paper, a new dynamic friction model for a pair of TSM will be presented with less model parameters in its structure and easy to implement for haptic problems. An illustration for NOTES system is introduced in Fig. 1.

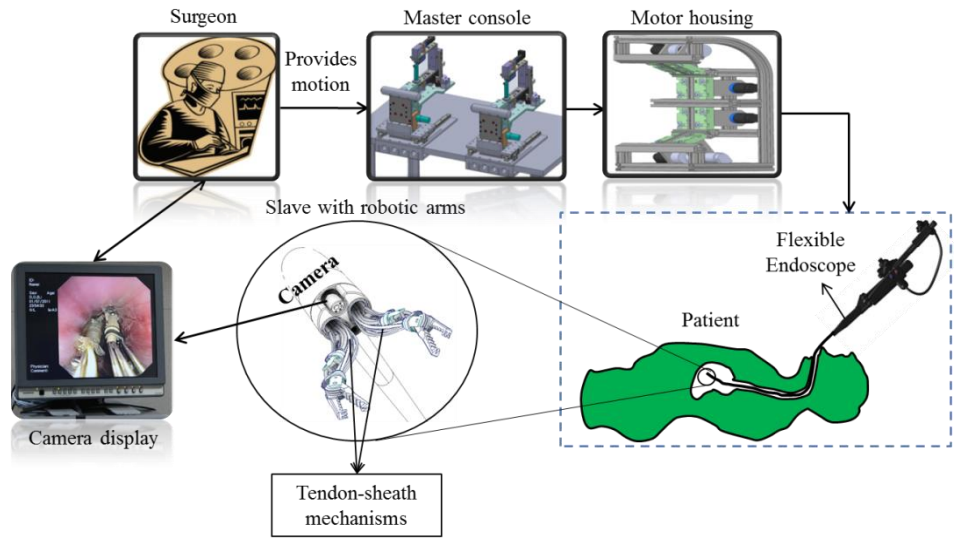

*Fig. 1*: Illustration of a Natural Orifice Transluminal Endoscopic Surgery (NOTES) system.

It is known that the normalized Bouc-Wen model can describe a wide range of hysteresis phenomena [36-38]. It is used in various applications like magnetorheological or piezoelectric systems. Therefore, a advanced Stribeck curve and a version of the modified normalized Bouc-Wen model will be developed to capture the nonlinearity characteristics of a pair of TSM. The ultimate goal of this approach is to provide the force information at the distal end of the endoscopic system without using any sensors at the distal end during operations. Having the force information, haptic devices can be utilized to provide necessary force to the surgeons. To validate the proposed approach, suitable experiments will be designed and comparisons will be given.

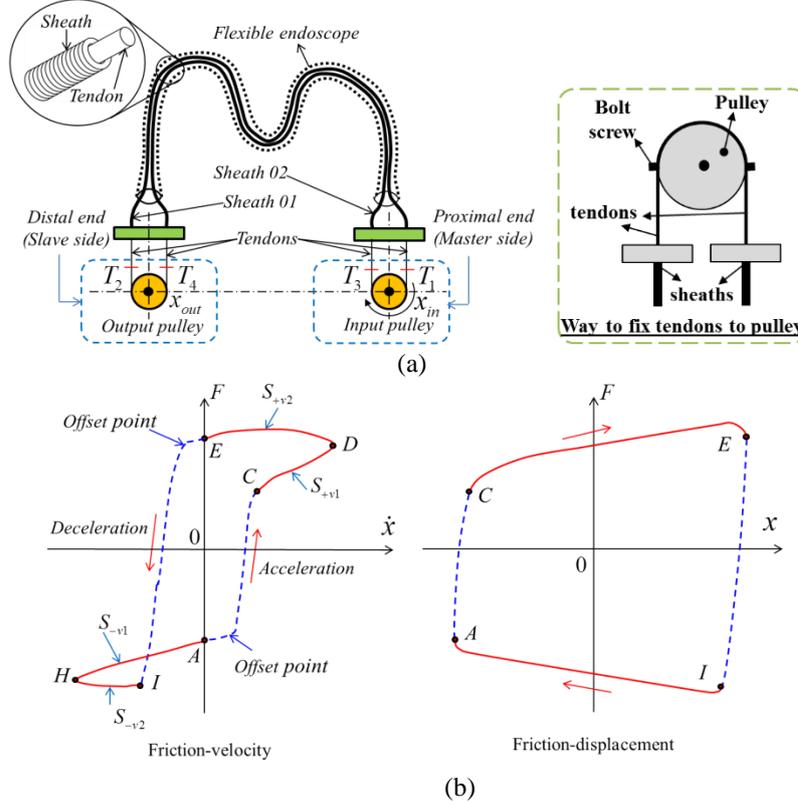

*Fig. 2*: A pair of tendon-sheath mechanisms
(a) (Left) Schematic of the TSMs, (Right) Ways to fix tendons on pulleys ; (b) Relation between friction force and motions: (Left) friction vs. velocity; (Right) friction vs. displacement

## 2. SYSTEM MODELLING

### 2.1 Mathematical methods

The diagram of a pair of TSMs using in flexible endoscopic systems is shown in Fig. 2. Let $x_{in}, x_{out}, \tau_{in}$, and $\tau_{out}$ denote the angular displacements and torques at the drive pulley (input pulley) and the follower pulley (output pulley), respectively. Let $T_i (i = 1, 2, 3, 4)$ denote the tensions at proximal end and distal end for each of the tendons. Assume that the two pulleys have the same radius $R$. The torques at the proximal end and distal end are expressed by:

$$\begin{cases} \tau_{in} = R(T_1 - T_3) \\ \tau_{out} = R(T_2 - T_4) \end{cases} \quad (1)$$

If we denote the friction forces in the Sheath 01 and 02 as $F_1$ and $F_2$, respectively, then the friction forces in term of tensions at proximal and distal end are given by:

$$\begin{cases} F_1 = T_1 - T_2 \\ F_2 = T_4 - T_3 \end{cases} \quad (2)$$

The total friction forces in the two sheaths can be calculated as follow:

$$F = \frac{\tau_{in} - \tau_{out}}{R} = F_1 + F_2 \tag{3}$$

The dynamic transmission profiles of a pair of TSMs are shown in Fig. 2(b) (the relation between total friction force and relative velocity of proximal end). Suppose that input pulley firstly rotates in the clockwise direction (positive velocity) as shown in Fig. 2(a). The motion at this pulley cannot be immediately propagated to the output pulley when the input pulley reverses its motion direction (to negative velocity). As a result, it exhibits some delays in response (points E to I in Fig. 2(b)). The friction force and the gap between the tendons and the sheaths prevent the immediate transmission of both motion and tension from the input pulley to the output pulley. As the friction on Sheath 01 decreases further, the output pulley begins to move–motion has been transmitted. When the friction in the Sheath 02 is further decreased, the two tendons in the Sheath 01 and Sheath 02 move together. Then, the output pulley follows immediately the input pulley (see points I to A in the Fig. 2(b)). Similar characteristics are exhibited for the reverse motion in the counter clockwise direction of a pair of TSMs. They are represented by points A to C and to D. It was also known that the pretension of the tendon affects the friction force in the tendon-sheath system [15, 23]; the pretension for the Tendon 01 inside the Sheath 01 is different with the one in the Sheath 02. In addition, the tension transmission across the two tendons is unequal. As a result, the friction forces in the Sheath 01 and 02 are different. Therefore, the total friction force $F$ is asymmetric for both negative and positive velocity. It can be observed that the friction characteristics for a pair of TSMs are similar to the one for a single TSM although there is a narrow gap in the hysteresis loops of the sliding regime (see Fig. 2(b)). This means that the previous dynamic friction models given in [31-33] can be used to characterize the nonlinearities of friction force for a pair of TSMs with a little change of the sliding curves. For validation and comparison, a novel friction model will be developed to characterize the dynamic nonlinearity of friction force for a pair of TSMs. Detailed experimental setup and results will be discussed in the next sections.

In order to capture the friction force for a pair of TSM, a first order derivative of internal state and advanced sliding curves are developed and they can be expressed by:

$$\dot{\zeta}(t) = \vartheta g \left( 1 - \left| \frac{\zeta}{F_{fnew}} \right|^n + \sigma \left| \frac{\zeta}{F_{fnew}} \right|^{n-1} \left( sgn\left( \frac{\zeta}{F_{fnew}} \right) - sgn(g) \right) \frac{\zeta}{F_{fnew}} \right) \tag{4}$$

where $g = (\dot{x} + \lambda sgn(\ddot{x}))$ is a function of acceleration $\ddot{x}$ and offset point $\lambda$ which can capture the offset points of hysteresis curves at vicinity of zero velocity for acceleration and deceleration direction.

$$F_{fnew}(\dot{x}(t), \ddot{x}(t)) = \begin{cases} S_{+v1} & \text{if } \dot{x} \geq 0, \ddot{x} \geq 0 \\ S_{+v2} & \text{if } \dot{x} \geq 0, \ddot{x} \leq 0 \\ S_{-v} & \text{if } \dot{x} \leq 0 \end{cases} \tag{5}$$

with

$$S_{+v1} = \rho_1 + \mu_1 e^{-\kappa_1 |\dot{x}|} + \mu_2 \left( 1 - e^{-|\kappa_2 \ddot{x}|} \right) \tag{6}$$

$$S_{+v2} = \rho_1 + \mu_1 e^{-\kappa_1 |\dot{x}|} - \mu_2 \left( 1 - e^{-|\kappa_2 \ddot{x}|} \right) \tag{7}$$

$$S_{-v} = \rho_2 + \mu_1 e^{-\kappa_1 |\dot{x}|} + sgn(\ddot{x}) \mu_3 e^{-f_2(\dot{x}, \ddot{x})} = \begin{cases} S_{-v1} & \text{if } \dot{x} \leq 0, \ddot{x} < 0 \\ S_{-v0} & \text{if } \dot{x} \leq 0, \ddot{x} = 0 \\ S_{-v2} & \text{if } \dot{x} \leq 0, \ddot{x} > 0 \end{cases} \tag{8}$$

The function $S_{-v}$ contains three sub-functions, which are expressed by:

$$S_{-v1} = \rho_2 + \mu_1 e^{-\kappa_1|\dot{x}|} + \mu_3 e^{-f_2(\dot{x},\ddot{x})} \text{ with } f_2(\dot{x},\ddot{x}) = \frac{\kappa_3|\dot{x}|}{|\ddot{x}|+\kappa_4} \quad (9)$$

$$S_{-v2} = \rho_2 + \mu_1 e^{-\kappa_1|\dot{x}|} - \mu_3 e^{-f_2(\dot{x},\ddot{x})} \quad (10)$$

$$S_{-v0} = \rho_2 + \mu_1 e^{-\kappa_1|\dot{x}|} \quad (11)$$

The new model for capturing the nonlinear friction force for a pair of TSMs can be rewritten by:

$$F(t) = \gamma \zeta(t) \quad (12)$$

where $f_1(\dot{x},\ddot{x}), f_2(\dot{x},\ddot{x})$ are velocity and acceleration dependent functions; $\rho_i, \mu_j, \kappa_l$ ($i=1,2$; $j=1,...,3$; $l=1,...,4$) are coefficients that control the shape of hysteresis loops in acceleration and deceleration direction.

### 2.2 Experimental Design of the Two-DOFs Master-Slave System

To validate the proposed approach, a dedicated experimental setup of the Two-DOFs Master-Slave system is presented. The system consists of a master console, telesurgical workstation (includes actuator housing and dSPACE controller), and a slave manipulator. For illustration, a slave system with two DOFs of robotic arm is introduced. The NOTES system has been illustrated in Fig. 1.

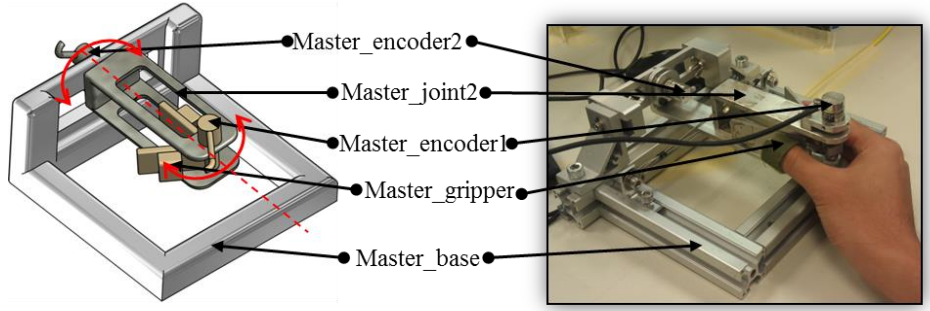

*Fig. 3*: Master console with two DOFs: (Left) Diagram; (Right) Photo.

The mechanism design of the master console with two DOFs is shown in Fig. 3. The master console, which allows the user to control the robotic arm at the distal end, is an ergonomic human-machine interface. In the experimental validation, a robotic arm with two rotational DOFs is used to validate the control schemes. A pair of TSMs is used to actuate the robotic joint. In the master console, encoders (Master_encoder1 and Master_encoder2) are mounted on the Master_gripper and Master_joint2, respectively; to provide necessary signals (position-reference trajectory $y_{r1}, y_{r2}$) to control the rotational joints (Joint1 and Joint2) with an attached gripper. The encoders are SCA16 from SCANCON. Signal from the Master_encoder1 and the Master_encoder2 will be subsequently sent to the dSPCAE controller where the data are processed and the tendon-sheath actuations are controlled. The users control the motions of the Joint1 and Joint2 via the Master_gripper and the Master_joint2, respectively. The picture of the master console is shown in the right panel of the Fig. 3.

The telesurgical workstation (Fig. 4) consists of an actuator housing and dSPACE DS1103 controller which is programmed via MATLAB Simulink from MathWorks. The signals from the master console and the robotic arm at the distal end are also acquired by this system. The actuator housing includes two Faulhaber 2232U024SR DC motors equipped with high resolution encoders IEH2. The slave joints (Joint1 and Joint2) are actuated by the two DC-motors via the TSMs. It is noted that each robotic joint is independently driven by a pair of the TSMs. The length of the two sheaths is 1.5 metre and they are routed along a flexible endoscope. To measure the tensions at the proximal end of the system, two load cells LW-1020-50 from Interface Corporation which are mounted on frictionless sliders are used.

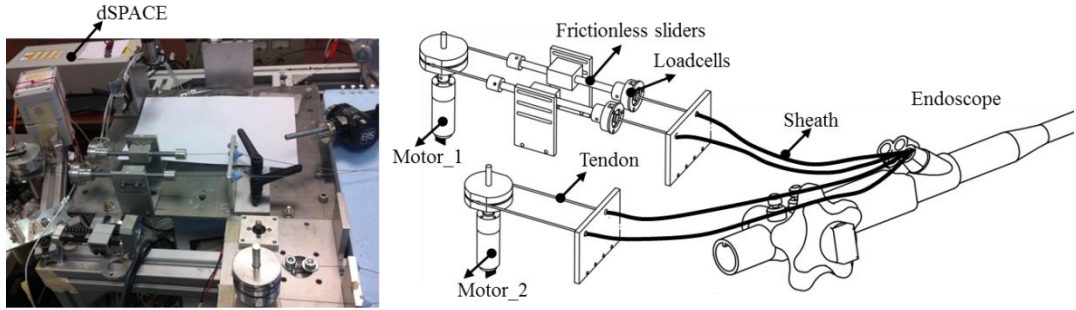

*Fig. 4*: Actuator housing with motor, encoder, and loadcells: (Left) Photo; (Right) Diagram.

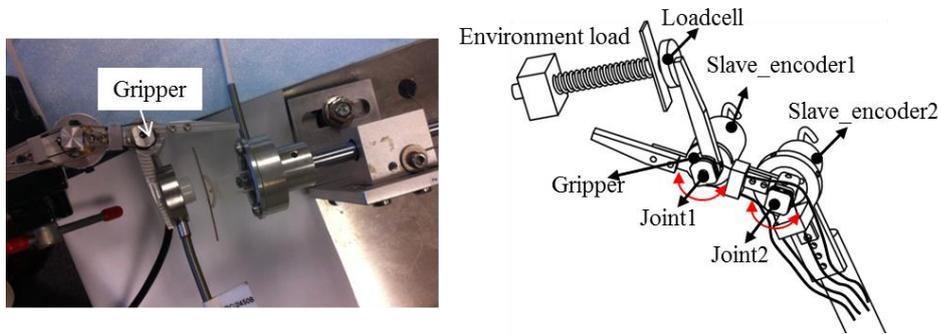

*Fig. 5*: Slave manipulator with two DOFs of joint and gripper: (Left) Photo; (Right) Diagram.

The slave manipulator structure is shown in Fig. 5. In the experimental work, the two DOFs of robotic arm are used for validation of the proposed schemes. A flexible endoscope, which is a type of GIF-2T160 from Olympus, Japan, has two tool channels with a length of 135cm for the flexible part. The robotic arm is mounted on the endoscope tip. At each side (proximal end and distal end), the tendons are attached to corresponding pulleys and joints. The Joint2 is actuated by the Motor_2 while the gripper (Joint1) is driven by the Motor_1. In the experimental work, a load cell LW-1020-50 from Interface Corporation which is attached to the gripper tip (see Fig. 6.4) to measure the interaction force between the gripper and the environmental load. High resolution encoders SCA16 from SCANCON are also used to measure the angular displacements of the joints. The tendons are connected to corresponding joints using bolt screws as presented by the Fig. 2(a).

### 2.3 Haptic feedback structure for the Two-DOFs Master-Slave system

The structure of the nonlinear adaptive control scheme with the presence of output angular position feedback is described in Fig. 6. In this paper, the compensation control scheme is based on the nonlinear adaptive control given in our previous works [24-27]. Let $R_i$ and $R_o$ are the radii of the input pulley at the proximal end of the Motor_1 and the output pulley at the distal end of the gripper joint (Joint1), respectively. The error positions between the proximal joints and distal joints are fed back to the control scheme. Subsequently, the proposed adaptive laws given by our previous works [24-27, 39, 40] are used to generate the control input $u_1, u_2$. Force data are measured only at the proximal end while the torque at distal end is estimated by the proposed dynamic friction model given in previous section. Hence, both accurate tracking performances and force information at the distal end can be achieved simultaneously.

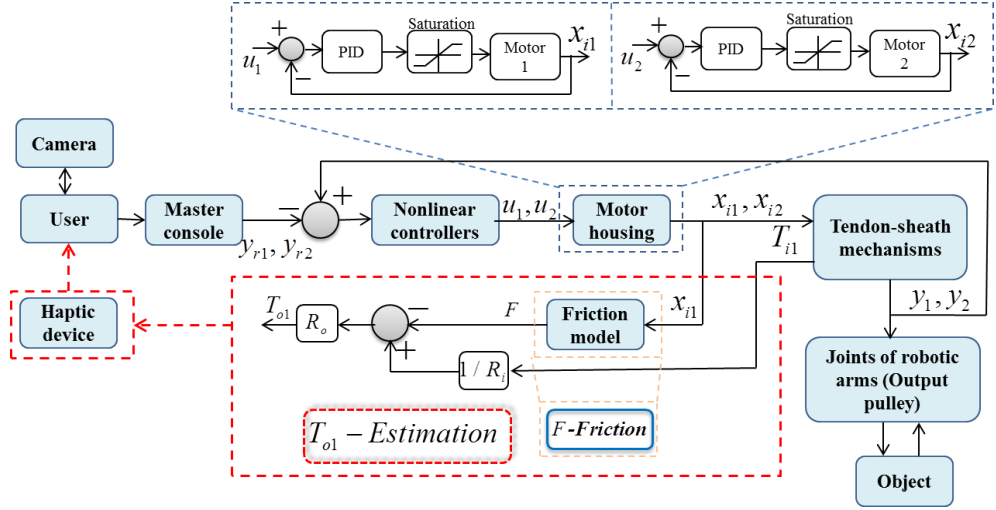

*Fig. 6*: Proposed compensation structures and haptic feedback
(With obtaining force feedback information for the Master-Slave system)

### 2.4 Validation results on the proposed Two-DOFs Master-Slave system

In this section, tracking performances between the desired trajectory and the output position will be revaluated using nonlinear control approach as presented in Fig. 6. The friction model structure which has been given by Fig. 7 will be utilized to provide accurate estimation of the output torque at the distal end of the gripper joint. Genetic algorithm is used to identify the friction model parameter. The obtained friction parameters are $\rho_1 = 1.038$, $\mu_1 = 1.0978$, $\kappa_1 = 6.1264$, $\mu_2 = 0.0458$, $\kappa_2 = 0.0269$, $\rho_2 = 0.8904$, $\mu_3 = 0.145$, $\kappa_3 = 13.248$, $\kappa_4 = 6.0297$, $\gamma = 9.5651$, $\lambda = 0.2$, $\vartheta = 15.5181$, $n = 2$, and $\sigma = 1$. Under the nonlinear adaptive control scheme and the availability of output position feedback, the backlash hysteresis phenomena are treated as uncertainties and the model parameters will be online estimated. See our previous works [24-27] for more details.

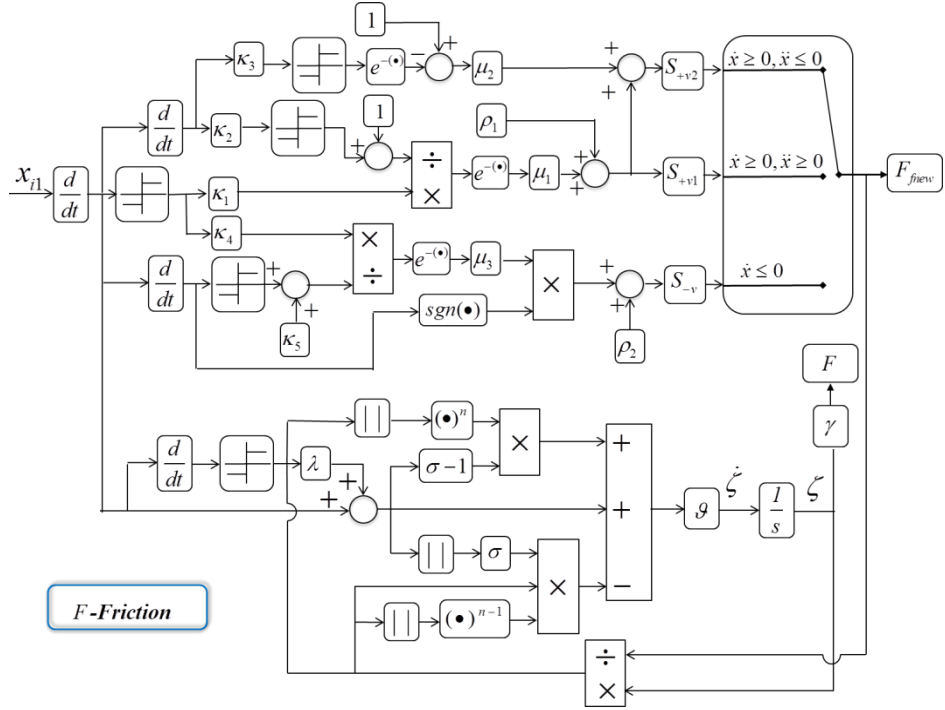

*Fig. 7*: Estimation of the friction force using the force information at the proximal end only

Consistent with previous trials, five experimental validations were carried out. For illustration purposes, one of the five trials will be given. Fig. 8 introduces the compensation results using the nonlinear adaptive control scheme of Fig. 6 for the Joint1. In the case of no compensation control, the measured position output $y_1$ always lags the desired trajectory $y_{r1}$. When the nonlinear control scheme is used (see Fig. 6), the measured output $y_1$ accurately follows the desired trajectory $y_{r1}$ (see the left panel of Fig. 8). The relative error under the nonlinear control scheme is also depicted in the right panel of Fig. 8. There is a significant reduction from 0.2669 rad peak-to-peak error before compensation to 0.08743 rad peak-to-peak after compensation.

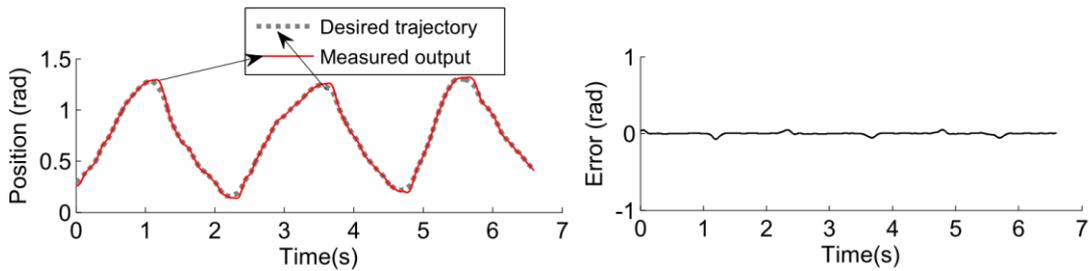

*Fig. 8*: Compensation results for one of the trials with nonlinear adaptive control scheme for Joint1 (Left) Time history of desired trajectory and measured displacement output after compensations; (Right) Tracking error.

The nonlinear and adaptive control scheme is also validated for Joint2. The left panel of Fig. 9 shows the time history of desired trajectory $y_{r2}$ and the measured output $y_2$ while the right panel presents the corresponding error. There is a high error of 0.2823 rad peak-to-peak before compensation compared to 0.0646 rad peak-to-peak after compensation. The tracking performances for the Joint2 can be observed from the right panel of Fig. 9.

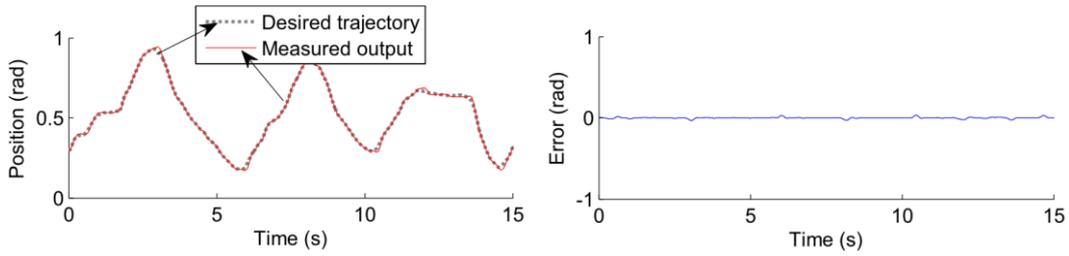

*Fig. 9:* Compensation results for one of the trials with nonlinear adaptive control scheme for Joint2 (Left) Time history of desired trajectory and measured displacement output after compensations; (Right) Tracking error.

The estimated friction force for the Joint1 is presented in the left panel of Fig. 10. There are small values of error (peak-to-peak of 2.6233 N) between the estimated friction and measured data. The time history of this error can be seen from the left panel of Fig. 10. The estimated output torque at distal end is also presented in the right panel of Fig. 10. From these results, it can be concluded that the desired tracking performances and estimated distal torque are achieved simultaneously. This results a good agreement between the theoretical model approaches and real-time implementation in a real Master-Slave system.

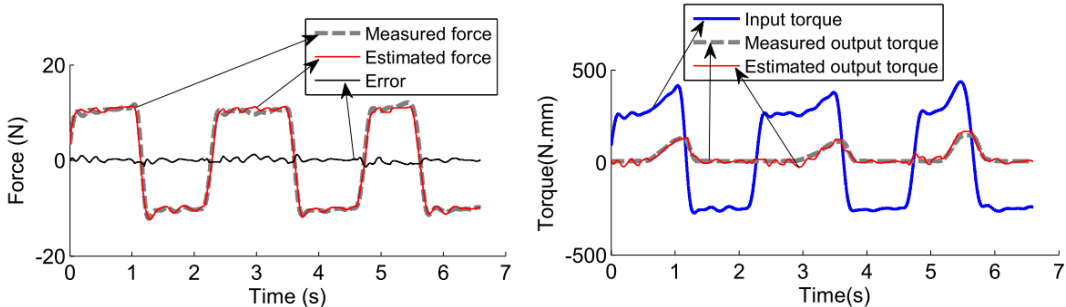

*Fig. 10:* Estimated torque of the Joint1 for the case of nonlinear adaptive compensation (Left panel) Time history of input torque, measured output torque, and estimated torque; (Right) Time history of tracking error result between the estimated output torque and measured output torque.

## 3. CONCLUSIONS

In this paper, the performances of the proposed controllers and friction estimation for a Two-DOF Master-Slave system have been presented. The proposed model and control schemes are experimentally investigated using a master console, a motor housing, and a slave manipulator with the two DOFs of robotic arm. The experimental results exhibit a good tracking performance between the robotic joints and the master console. It has been demonstrated that the proposed control schemes including a feedforward compensation and a nonlinear adaptive control works well on the real Master-Slave system. Desired tracking performances and force information are simultaneously achieved. It has also been shown that the proposed schemes are able to track the desire reference signals generated by the user. In addition, the results have also demonstrated that the dynamic friction models are able to provide accurate estimation of the force feedback information where only external sensors are used.

Although the control architecture and force feedback scheme have been demonstrated for the Two-DOF Master-Slave system, they can be used in any flexible endoscopic system with a master-slave manipulator with tendon-sheath-driven robotic arms. In the absence of the measured position feedback at the distal end, direct inverse hysteresis model-based feedforward scheme is a potential solution. If the output displacement is available for feedback, nonlinear adaptive control can be a better alternative to enhance the tracking performances with higher accuracy regardless of the change of sheath configuration. The proposed control scheme and force feedback structures used the position as the control input and estimated friction model and external sensors located at the proximal end as the force feedback. These structures will benefit any flexible endoscopic robots that use the TSMs as the main mode of transmission.